\documentclass{PoS}

\title{VRIJHK PHOTOMETRY OF PKS 0537-441 AND PKS 2155-304 IN 2006-2008}

\ShortTitle{VRIJHK photometry of PKS 0537 - 441 and PKS 2155 - 304}

\author{\speaker{D. Impiombato}\thanks{A footnote may follow.}\\
        Physics Dept., University of Perugia, Perugia, Italy\\
        E-mail: \email{Domenico.Impiombato@pg.infn.it}}

\author{Gino Tosti\\
        Physics Dept., University of Perugia, Perugia, Italy\\
        E-mail: \email{Gino.Tosti@pg.infn.it}}

\author{Aldo Treves\\
        Physics Dept., University of Insubria, Como Italy\\
        E-mail: \email{Aldo.Treves@mib.infn.it}}

\author {Stefano Covino\\
        INAF/AO Brera, Merate, Italy \\
        E-mail: \email{stefano.covino@brera.inaf.it}}

\author {Elena Pian\\
        INAF/AO Trieste, Italy\\
        E-mail: \email{pian@oats.inaf.it}}
        
\author {Stefano Ciprini\\
        Physics Dept., University of Perugia, Perugia, Italy\\
        I.N.F.N. Section of Perugia, via A. Pascoli, 06123 Perugia, Italy\\
        Tuorla Observatory, University of Turku,
        V\"{a}is\"{a}l\"{a}ntie 20, 21500 Piikki\"{o}, Finland\\
        E-mail: \email{stefano.ciprini@pg.infn.it}}

\abstract{We report the 2006-2008 light curves obtained with the REM telescope in VRIJHK bands for
the two BL Lac objects PKS 0537-441 and PKS 2155-304}

\FullConference{Workshop on Blazar Variability across the Electromagnetic Spectrum\\
		 April 22-25 2008\\
		 Palaiseau, France}

\begin{document}

\section{Introduction}
Since 2004 we are using the Rapid Eye Mount robotic telescope (REM, la Silla Chile, Zerbi et al 2004),
for an intensive Blazar monitoring program in near-infrared (JHK) and optical (IRV) bands (see Tosti, these proceedings).
The remarkable activities of PKS 0537- 441 (z=0.896), and PKS 2155-304 (z=0.116) observed with REM in 2004-2005
are described in Dolcini et al. (2005, 2007) and in Pian et al (2007).\\
Here we report preliminary results obtained for these two sources in 2006-2008.

 \section{PKS 0537-441}
 The light curves in the six bands are reported in fig.1. Typical  photometric errors are $\sim$ 5 $\textperthousand$ .\\
 The 2007 November and the December observations clearly indicate the decaying part of a large flare. 
 At its maximum the source was close to the peak of the flaring state detected in February 2005 and reported 
 by Dolcini et al.2005. The decay time  (~40 days) and the magnitude variation $\bigtriangleup$m $\approx$ 1.8 mag  are also similar.
 In February -March 2008 the source is in a brightening phase.\\ It increased by 2.3 mag in 17 days in the V band.\\
 Spectral energy distributions in four intensity states are reported in figure 2, with a power law best fit. The tendency of the source of
 becoming harder at higher fluxes reported by Dolcini et al. 2005, is clearly contrasted. \\
\indent
The February-March 2008 observing period is  simultaneous with an AGILE  exposure of  the source.\\
\section{PKS 2155 -304}
The light curves are given in fig.3.
The source appears very active in August $\div$ Nov 2006, reaching a maximum of V= 12.02 mag.
Note that at the end of July 2006 a large TeV flare was detected by Aharonian et al. 2007 
(see also Foschini et al. 2007 and 2008, Sakamoto et al. 2008), but  no REM data are available.\\
\indent
The observations 2006 November 4 are simultaneous with an X-ray spectrum taken with Newton XMM 
(Zhang et al. 2008, Foschini et al 2008), which may represent a transition from HBL to LBL behavior.
 No special features appear in the optical bands. \\
\indent
During our monitoring the source was variable in all bands with a excursion of $\bigtriangleup$m $\approx$1.2 mag.\\
In 2005, Dolcini et al. (2007) reported a flare lasting several days, which was more prominent in the infrared bands.
Such rather unexpected behavior does not appear in 2006-2007 (see the V-K light curve given in figure 4).\\

\vspace{2cm}

 \begin{figure*}[thb!!]
\centering
\includegraphics[angle =0, width=10cm, height=10cm]{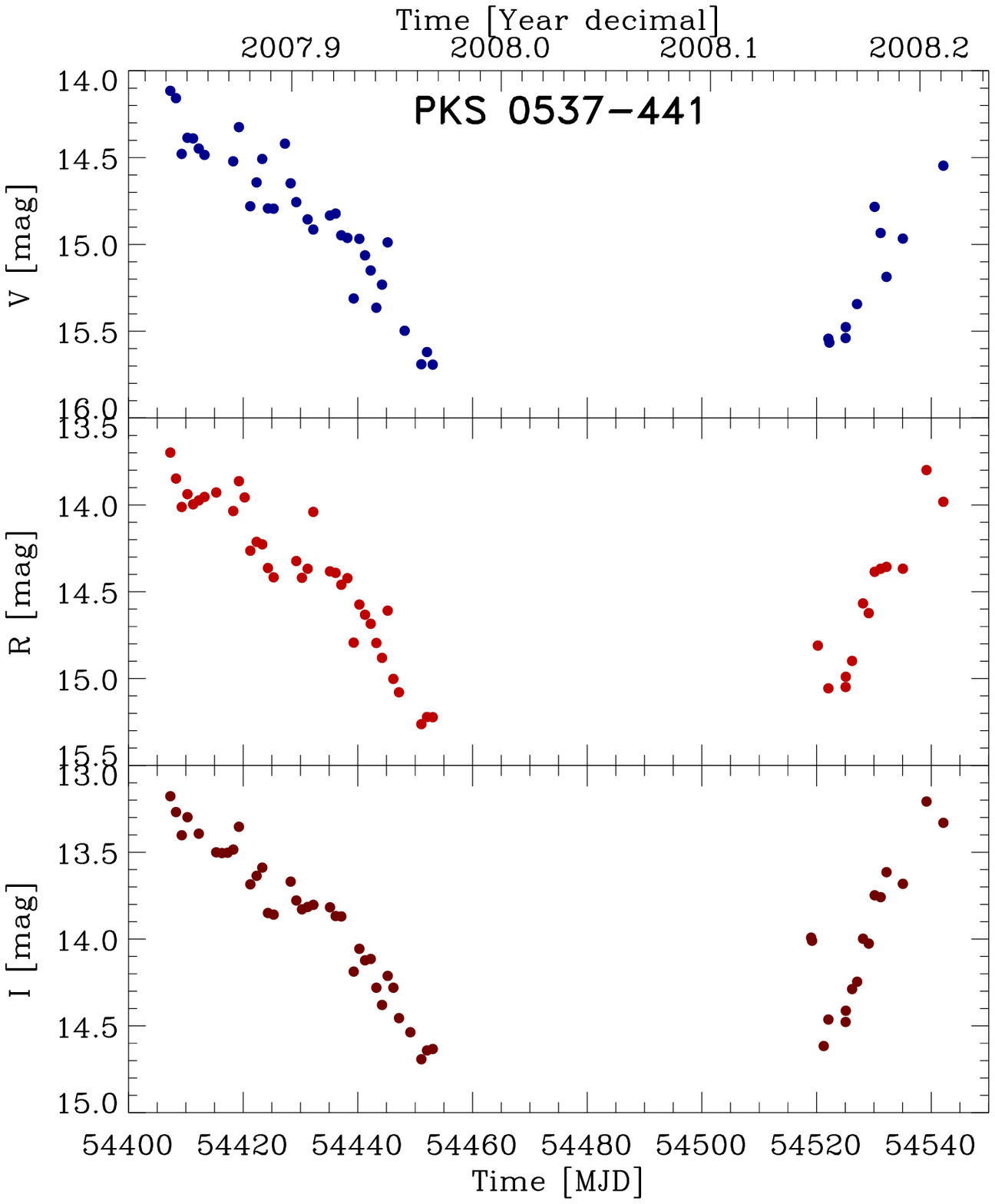}\qquad
\includegraphics[angle =0, width=10cm, height=10cm]{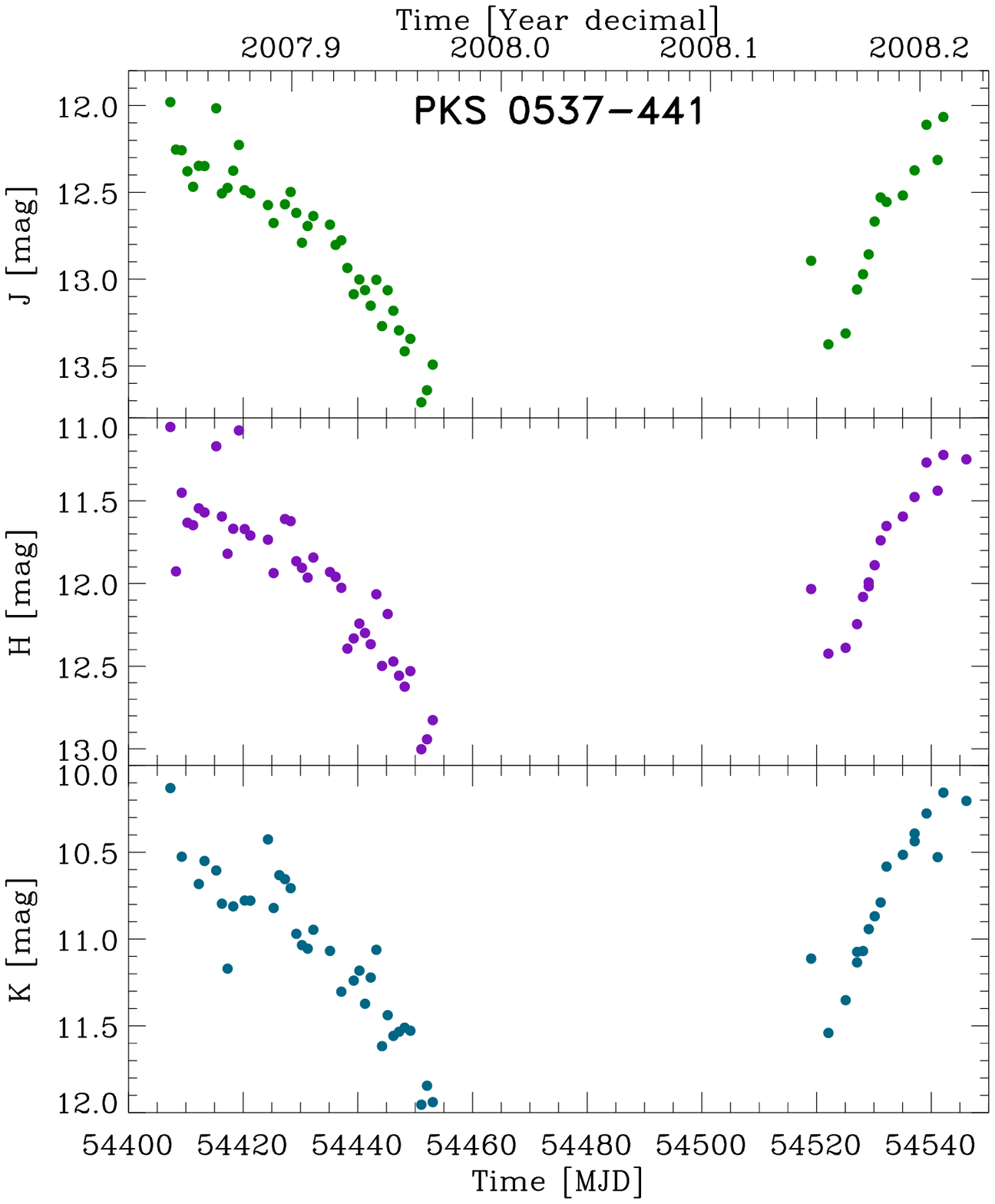}
\caption{ \footnotesize The REM VRIJHK light curves of PKS0537- 441 in 2006-2008. The typical error is $\sim$ 0.05 mag.\
}
 \label{figura2}
\end{figure*}

\begin{figure}[c]
\centering
\includegraphics[angle =90, width=12cm, height=12cm]{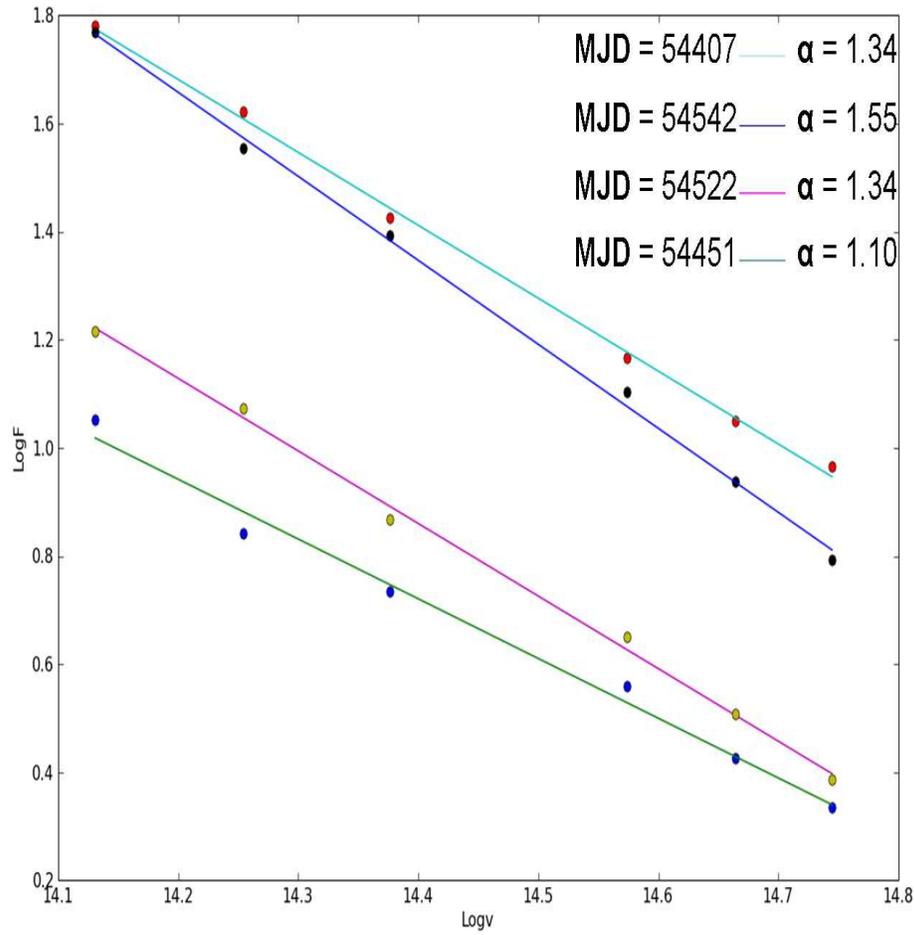}
\caption{ \footnotesize Spectral flux distribution of PKS 0537-441 at various epochs. Fluxes $F_{\nu}$ are in mJy, frequency $\nu$ in Hz.\
}
 \label{figura2}
\end{figure}

 \newpage

\begin{figure*}[htbp]
\centering
\includegraphics[angle =0, width=7cm, height=7cm]{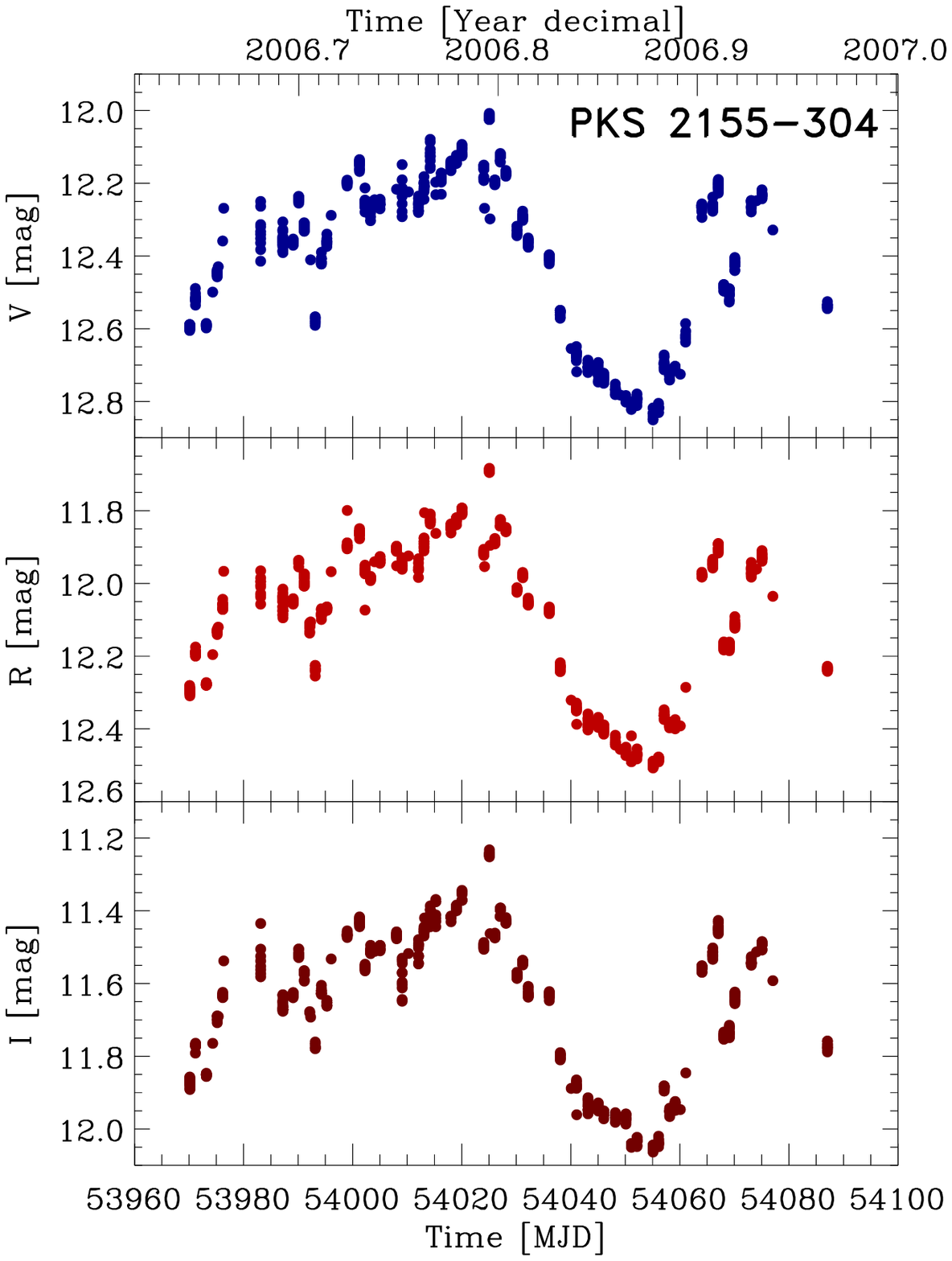}
\qquad
\includegraphics[angle =0, width=7cm, height=7cm]{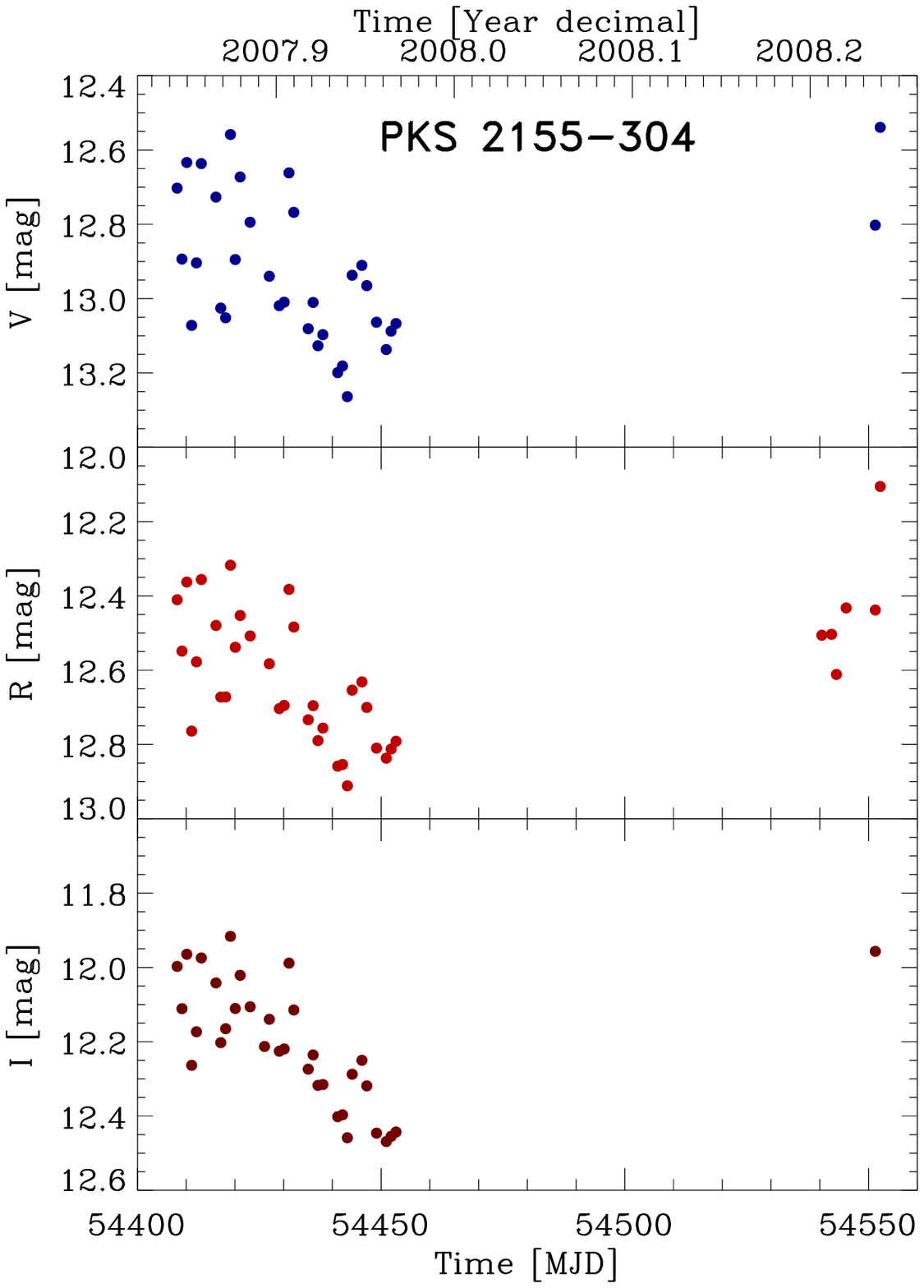}
\includegraphics[angle =0, width=7cm, height=7cm]{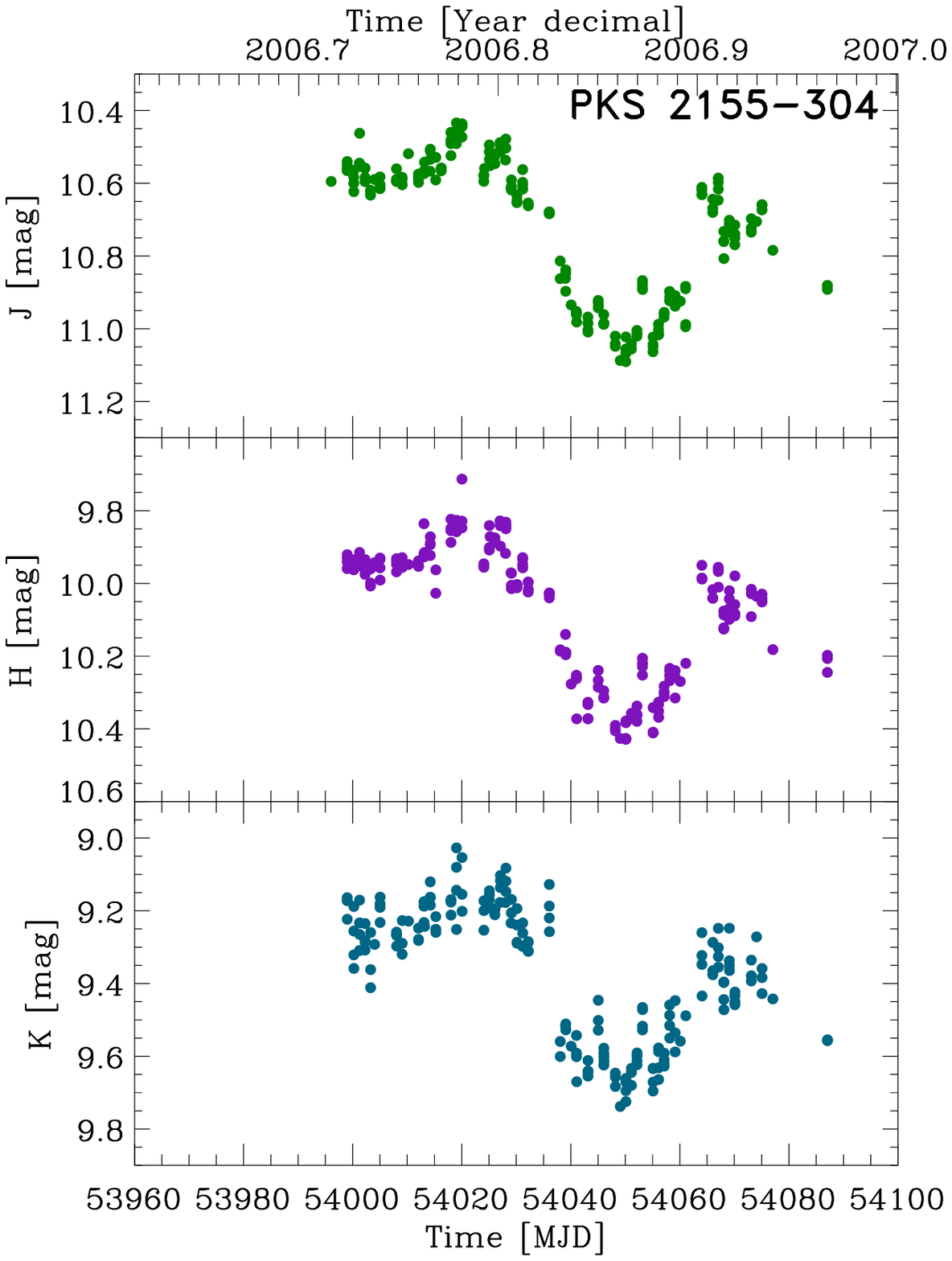}
\qquad
\includegraphics[angle =0, width=7cm, height=7cm]{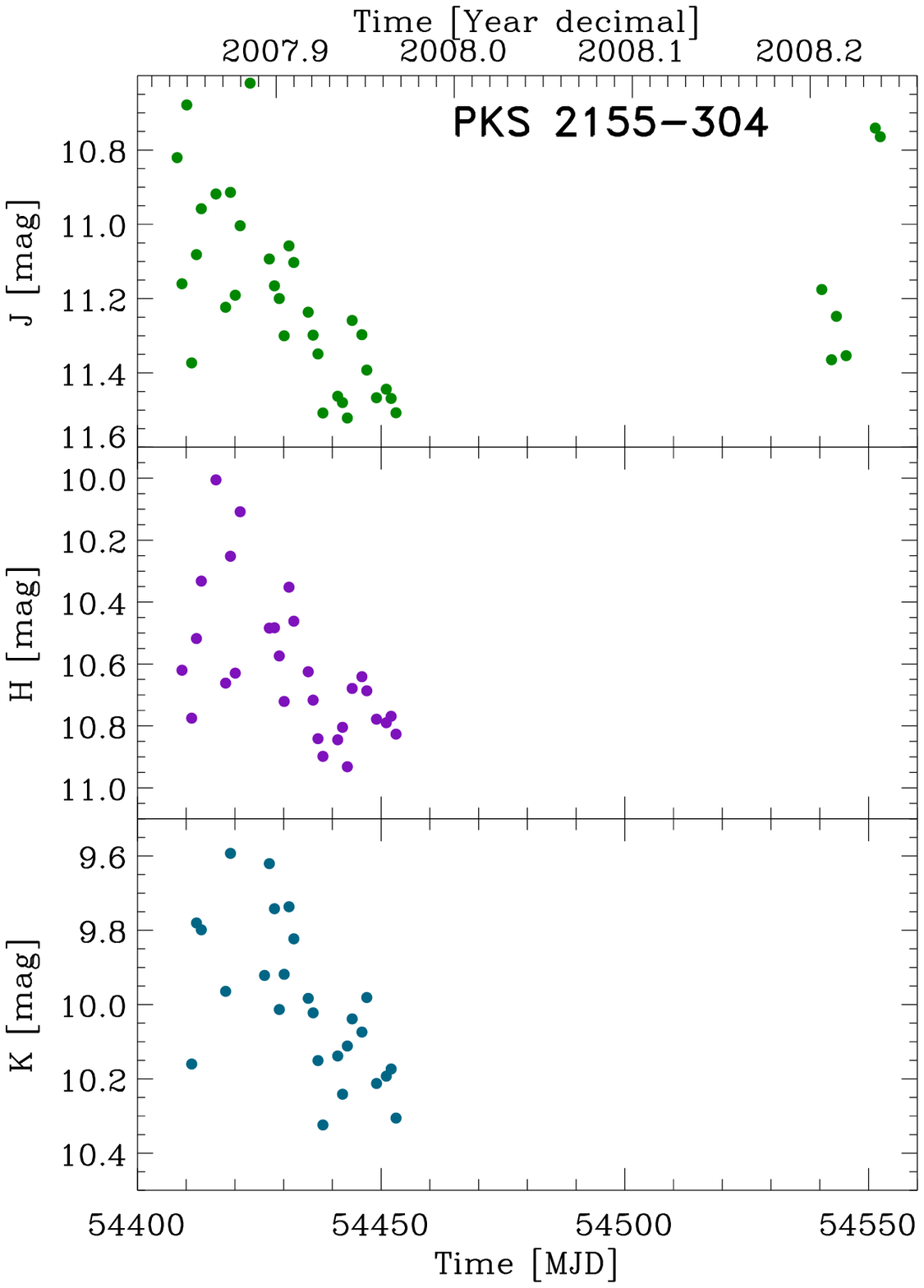}
\caption{ \footnotesize VRIJHK light curves of PKS2155-304. The typical error is $\sim$ 0.05 mag  }
 \label{figura}
 \end{figure*}

\newpage
\vspace{1.5cm}
\begin{figure*}[htbp]
\centering
\includegraphics[angle =90, width=7cm, height=7cm]{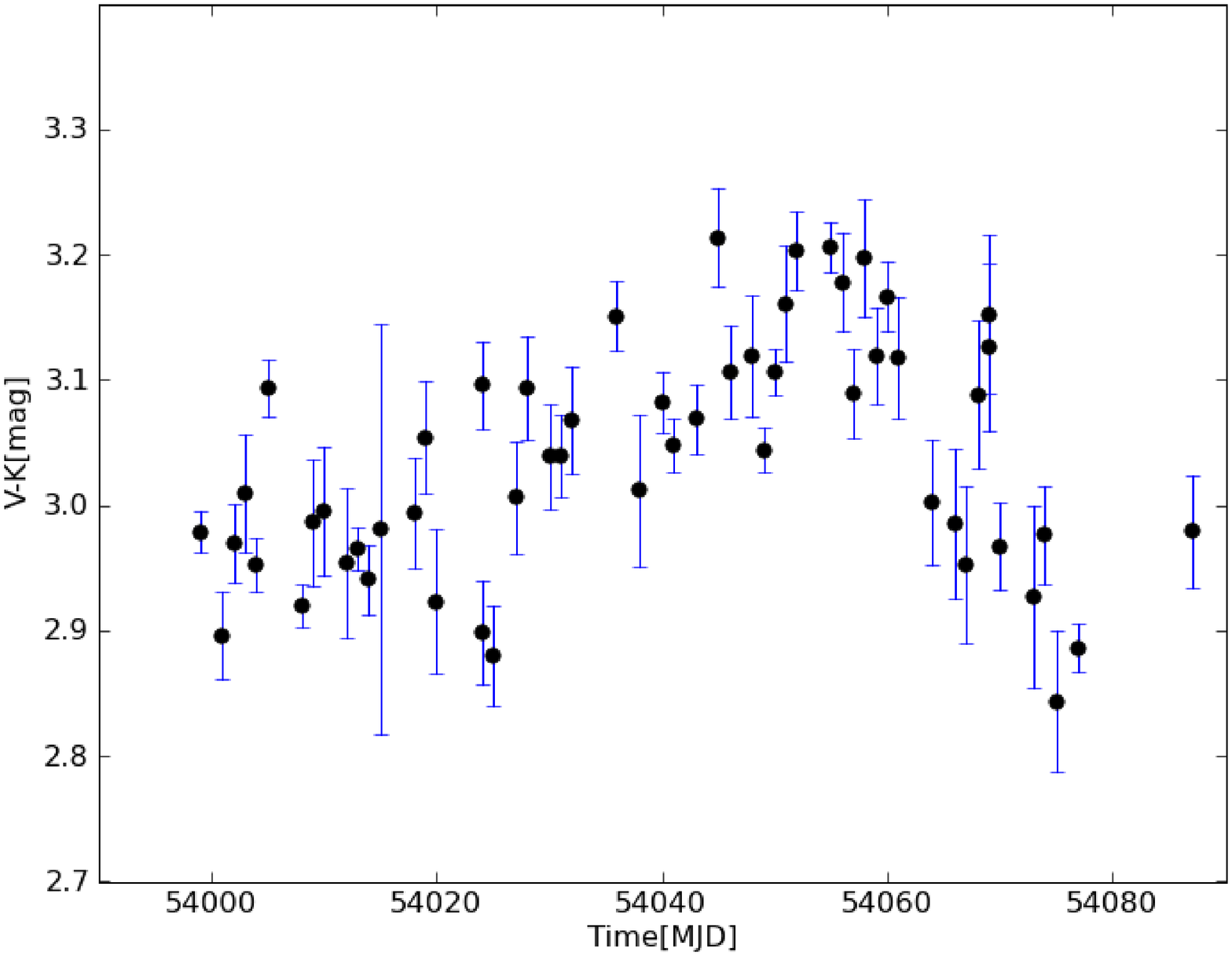}
\qquad
\includegraphics[angle =90, width=7cm, height=7cm]{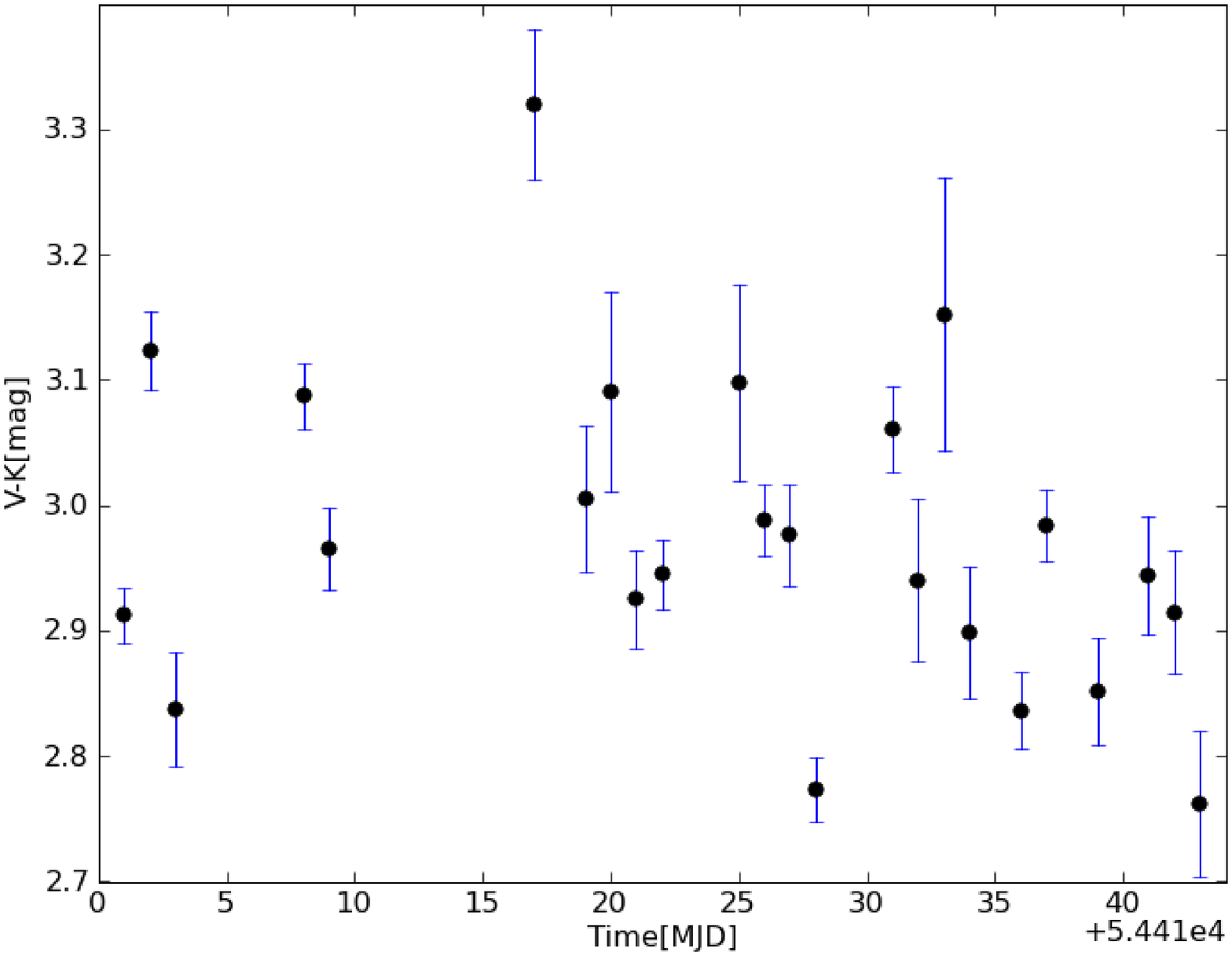}
\caption{ \footnotesize V-K colour index versus time. The colour magnitude errors are indicated by the vertical lines.}
\label{figura}
\end{figure*}

\newpage

\end{document}